\documentclass[fleqn,usenatbib]{mnras}
\usepackage{newtxtext,newtxmath}
\usepackage[T1]{fontenc}
\usepackage{graphicx}	
\usepackage{amsmath}	
\usepackage{CJKutf8}
\usepackage{array}
\usepackage[figuresright]{rotating}
\usepackage{tabularx}

\title[L222\_78: an outbursting YSO]{Multi-wavelength detection of an ongoing FUOr-type outburst on a low-mass YSO}

\author[Z. Guo et al.]{Zhen Guo$^{1, 2, 3, 4}$\thanks{E-mail: zhen.guo@uv.cl},
P. W. Lucas$^{4}$,  R. G. Kurtev$^{1,5}$, J. Borissova$^{1,5}$,  Vardan Elbakyan$^{6,7}$,  C. Morris$^{4}$, 
 \newauthor
 A. Bayo$^{8}$, L. Smith$^{9}$,
 A. Caratti o Garatti$^{10,11}$,  C. Contreras Pe{\~n}a$^{12, 13}$, D. Minniti$^{14, 15, 16}$, J. Jose$^{17}$,
  \newauthor
  M. Ashraf$^{17}$, J. Alonso-Garc\'{i}a$^{18,5}$, N. Miller$^{4}$, and H.  D. S. Muthu$^{4}$
\\
$^{1}$Instituto de F{\'i}sica y Astronom{\'i}a, Universidad de Valpara{\'i}so, ave. Gran Breta{\~n}a, 1111, Casilla 5030, Valpara{\'i}so, Chile\\
$^{2}$N\'ucleo Milenio de Formaci\'on Planetaria (NPF), ave. Gran Breta{\~n}a, 1111, Casilla 5030, Valpara{\'i}so, Chile\\
$^{3}$Departamento de F{\'i}sica, Universidad Tecnic{\'a} Federico Santa Mar{\'i}a, Avenida Espa{\~n}a 1680, Valpara{\'i}so, Chile\\
$^{4}$Centre for Astrophysics Research, University of Hertfordshire, Hatfield AL10 9AB, UK\\
$^{5}$Millennium Institute of Astrophysics, Nuncio Monse{\~n}or Sotero Sanz 100, Of. 104, Providencia, Santiago, Chile\\
$^{6}$Fakultät für Physik, Universität Duisburg-Essen, Lotharstraße 1, D-47057 Duisburg, Germany\\
$^{7}$Research Institute of Physics, Southern Federal University, Stachki Ave. 194, Rostov-on-Don 344090, Russia\\
$^{8}$ European Organisation for Astronomical Research in the Southern Hemisphere (ESO), Karl-Schwarzschild-Str. 2,
85748 Garching bei München, Germany\\
$^{9}$Institute of Astronomy, University of Cambridge, Madingley Road, Cambridge, CB3 0HA, UK\\
$^{10}$INAF - Osservatorio Astronomico di Capodimonte, salita Moiariello 16, 80131, Napoli, Italy\\
$^{11}$Dublin Institute for Advanced Studies, School of Cosmic Physics, Astronomy and Astrophysics Section, 31 Fitzwilliam Place, Dublin 2, Ireland\\
$^{12}$Department of Physics and Astronomy, Seoul National University, 1 Gwanak-ro, Gwanak-gu, Seoul 08826, Republic of Korea\\
$^{13}$Research Institute of Basic Sciences, Seoul National University, Seoul 08826, Republic of Korea\\
$^{14}$Departamento de Ciencias Fisicas, Universidad Andres Bello, Republica 220, 8320000 Santiago, Chile\\
$^{15}$Vatican Observatory, V00120 Vatican City State, Italy\\
$^{16}$Departamento de F{\'i}sica, Universidade Federal de Santa Catarina, Trindade 88040-900, Florianopol{\'i}s, SC, Brazil\\
$^{17}$Indian Institute of Science Education and Research (IISER) Tirupati, Rami Reddy Nagar, Karakambadi Road, Mangalam (PO), Tirupati 517 507, India\\
$^{18}$Centro de Astronom{\'i}a (CITEVA), Universidad de Antofagasta, Av. Angamos 601, 02800 Antofagasta, Chile\\
}

\date{Accepted XXX. Received YYY; in original form ZZZ}
\pubyear{2023}

\begin{document}
\label{firstpage}
\pagerange{\pageref{firstpage}--\pageref{lastpage}}
\maketitle

\begin{abstract}
During the pre-main-sequence evolution, Young Stellar Objects (YSOs) assemble most of their mass during the episodic accretion process. The rarely seen FUOr-type events (FUOrs) are valuable laboratories to investigate the outbursting nature of YSOs. Here, we present multi-wavelength detection of a {  high-amplitude eruptive source} in the young open cluster VdBH~221 with an ongoing outburst, including optical to mid-infrared time series and near-infrared spectra. The initial outburst has an exceptional amplitude of $>$6.3~mag in {\it Gaia} and 4.6~mag in $K_s$, with a peak luminosity up to 16 $L_{\odot}$ and a peak mass accretion rate of  1.4 $\times$ 10$^{-5}$ $M_\odot$ yr$^{-1}$. The optical to infrared spectral energy distribution (SED) of this object is consistent with a low-mass star (0.2$M_\odot$) with a modest extinction ($A_V < 2$ mag). A 100-d delay between optical and infrared rising stages is detected, suggesting an outside-in origin of the instability. The spectroscopic features of this object reveal a self-luminous accretion disc, very similar to FU~Orionis, with a low line-of-sight extinction. Most recently, there has been a gradual increase in brightness throughout the wavelength range, possibly suggesting an enhancement of the mass accretion rate.
\end{abstract}




\section{Introduction}

A FUOr-type event, named after the prototype FU~Orionis, displays a prominent outburst in its light curve, usually with an amplitude larger than 5~mag in the optical \citep[see the review from][]{Fischer2022}. These events are rarely detected, as once per 10$^{4}$ to 10$^{5}$ yrs per star, estimated by the total known eruptive samples \citep[][and Contreras Pe\~na submitted]{Scholz2012, Hillenbrand2015, Contreras2019}. So far, only two dozen such events have been confirmed by photometric and spectroscopic data on young stars. Most FUOr-type outbursts share two common photometric signatures \citep{Hartmann1996}, high amplitude ($\sim$5 mag in optical) and long duration (tens to nearly one hundred years). These two characteristics distinguish FUOrs from the low-amplitude (up to a few mags) and short-timescale (up to a few years) events, such as EXOr-type outbursts with magnetospheric accretion \citep{Herbig2007}.  More recently, the eruptive phenomenon has been observed at all stages and stellar masses in star formation, especially among embedded protostars \citep{Contreras2017}. Notably, most infrared-detected eruptive YSOs have intermediate observational features between the classical FUOr and EXOr groups \citep[see][]{Contreras2017b}. Theoretical models have been proposed to trigger episodic accretion bursts on YSOs, including gravitational instabilities \citep[GI,][]{Armitage2001, Kratter2016}, thermal instability \citep{Lodato2004, Clarke2005}, magneto-rotational instabilities \citep[MRI, e.g.][]{Zhu2009b, Elbakyan2021}, disc fragmentation \citep[e.g.][]{Vorobyov2015}, stellar fly-by \citep {Borchert2022} and evaporation of gas giant planets \citep{Nayakshin2023}. 

During the FUOr event, unlike the steady magnetospheric accretion seen on most disc-bearing YSOs, the disc material is directly accreted onto the star by the boundary layer accretion mode \citep[][]{Audard2014}. The mass accretion rate during a solar mass FUOr-type outburst can reach 10$^{-4} M_\odot \rm yr^{-1}$, orders of magnitudes higher than the steady accretion stage. Episodic accretion models predict that most of the stellar mass is accumulated during these outbursts \citep{Hartmann1996}. The released gravitational energy can efficiently heat the inner accretion disc, which becomes self-luminous and outshines the photospheric emission \citep{Zhu2009, Liu2022}. The near-infrared (NIR) spectra of FUOrs resemble a bright but cool object, with strong absorption bands mainly from the molecules in the circumstellar disc/envelope. Most FUOrs are confirmed by a combination of eruptive photometric light curves along with their unique spectral features \citep{Connelley2018, Guo2021}.

The eruptive behaviours of FUOrs across different wavelength ranges can provide clues on the origin of the triggering instability \citep[][]{Vorobyov2021}. For instance, Gaia17bpi had a pre-outburst in the mid-infrared, 500 days before the optical outburst \citep[][]{Hillenbrand2018}, indicating an outside-in propagation of the MRI originated at $\sim$1~AU \citep{Cleaver2023}. In contrast, slow-rising outbursts in both optical and infrared bands are explained by inside-out propagating instabilities \citep[e.g.][]{Lin1985}. Theoretical models predict months-long delays in the mid-IR light curves when the outburst is triggered by thermal instabilities initiated at several stellar radii from the star \citep{NayakshinElbakyan2023}. The fading stage of FUOrs contains information about the cooling efficiency of the viscous heated accretion disc. The prototypes of FUOrs have remained bright for decades, though fading at quite different rates \citep{Hartmann1996}, but some recently discovered bona fide FUOrs have a rapid fading stage, which has blurred the initial classification of FUOrs. Previous studies preferred the enhancement of line-of-sight extinction as the explanation of the post-outburst rapid decays \citep[][]{Kopatskaya2013, Hackstein2015}. However, analytical disc models suggest that the rapid cooling of the accretion disc can result in such decay in the brightness \citep[][]{Szabo2021, Carvalho2023}. 

Lucas et al. (submitted, hereafter LSG23) discovered 222 high-amplitude objects ($\Delta K_s > 4$ mag) from the decade-long Vista Variables in the Via Lactea survey \citep[VVV,][]{Minniti2010, Saito2012, Minniti2016}. In this work, we present multi-wavelength light curves and spectra of an ongoing {  outbursting YSO} (L222\_78, RA: 17:18:19.65 Dec: -32:22:53.11) from the catalogue provided by LSG23. This target is associated with the young open cluster VdBH 221 \citep[][]{Cantat2020}. The outburst on this object reached $6.3$~mag in {\it Gaia} $G$-band, followed by a short fading stage and a unique recent brightening across the wavelength spectrum.

\section{Observation and data reduction}
\label{sec: data}
The eruptive behaviour of L222\_78 was originally discovered in LSG23 using the $K_s$ time series from the VVV Infrared Astrometric Catalogue \citep[VIRAC2$\beta$,][and in prep]{Smith2018}. The VIRAC2$\beta$ catalogue provides PSF (point-spread function) photometry of dozens of $K_s$ detections (from 2010 to 2019) and two multi-colour ($Z, Y, J, H$) epochs. A few epochs at/after the photometric maximum are saturated in the VVV images, which were corrected by custom-designed aperture photometry in LSG23. 

In 2021 and 2022, we obtained single-epoch $J, H, K_s$ photometry of this target using the Son of ISAAC infrared imager on the ESO NTT telescope \citep[SOFI][]{Moorwood1998}. We obtained optical to NIR photometry from the SMARTS 1 m telescope in 2022 and the Rapid Eye Mount (REM) telescope in 2023. Two epochs of $r$ and $i$-band images are found in the LCO science archive \citep[][]{Brown2013}. Plus, we retrieved $g$, $r$, and $i$-band images of the {\it VPHAS+} survey taken by the VLT Survey Telescope (VST) from the ESO archive \citep{Drew2014}, and broadband calibrated images from  {\it ATLAS} forced photometry server \citep{Tonry2018}. Custom-designed aperture photometry measurements were applied to extract the brightness \citep[see][]{Guo2018a}. 


We retrieved optical light curves and astrometry data from {\it Gaia} DR3 \citep{Gaia2016, Gaia2022}. Since this object is not included in the Gaia Photometric Science Alert, the archived data only includes observations up until circa 2017. The {\it Gaia} parallax measurement of L222\_78 is 0.9209$\pm$0.0218 mas, which is equivalent to a distance of 1.08$\pm$0.02 kpc. 

We obtained mid-infrared photometric data from {\it ALLWISE} \citep{Wright2010} and {\it NEOWISE} \citep{Mainzer2014} surveys via the NASA/IPAC Infrared Science Archive. Due to the low spatial resolution, L222\_78 is blended with a nearby companion {  ($d = 2.2"$)}. We performed custom-written PSF photometry on these two sources, using the epoch-binned {\it unWISE} images \citep{Lang2014a, Meisner2017}, obtained from the online {\it wiseview} tool\footnote{\url{http://byw.tools/wiseview}}. The PSF functions were generated in each image (2' x 2') by the {\sc daophot} package, using the 30 brightest sources in the field \citep{Stetson1987}. Then, the PSF function is applied to the target and companion with fixed central locations obtained from the VVV survey. Finally, we obtained the best-fit heights of the PSF functions for both sources using the least $\chi^2$ method. The typical uncertainty around the photometric maximum is 0.1~mag and enhanced to 0.5~mag around the photometric minimum. There was no detection from {\it Spitzer}.  

\begin{figure*}
\includegraphics[height=3.45in]{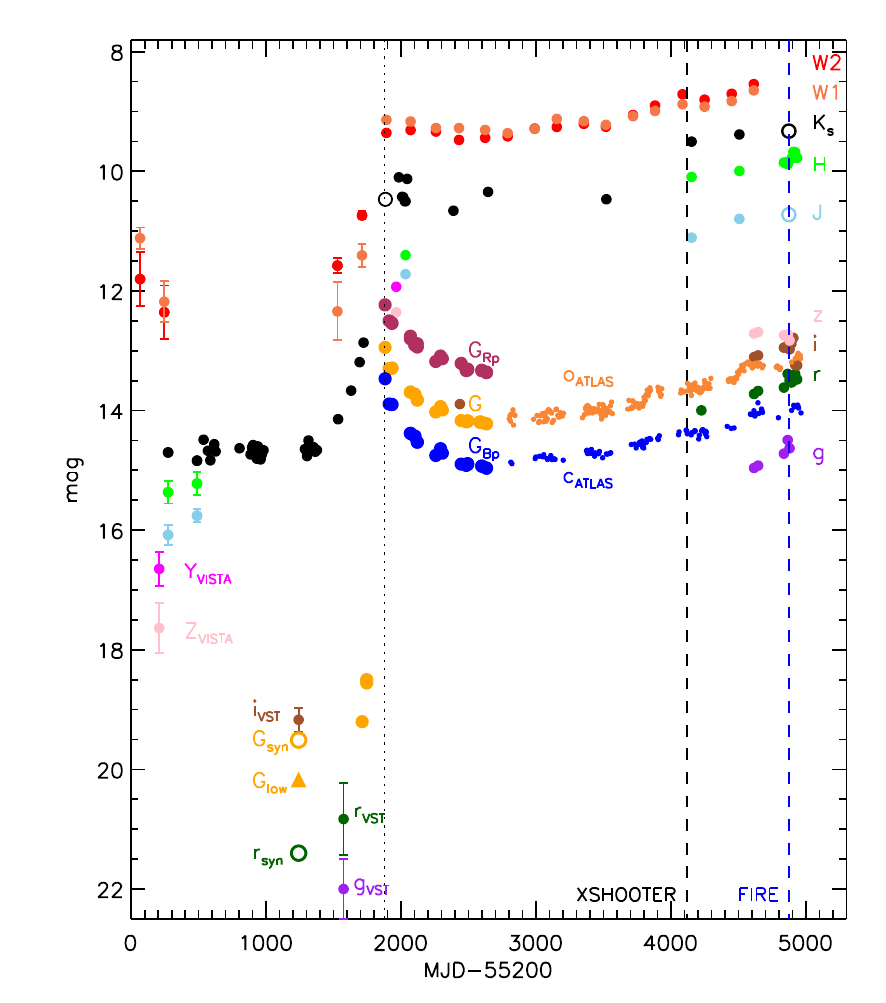}
\includegraphics[height=3.4in]{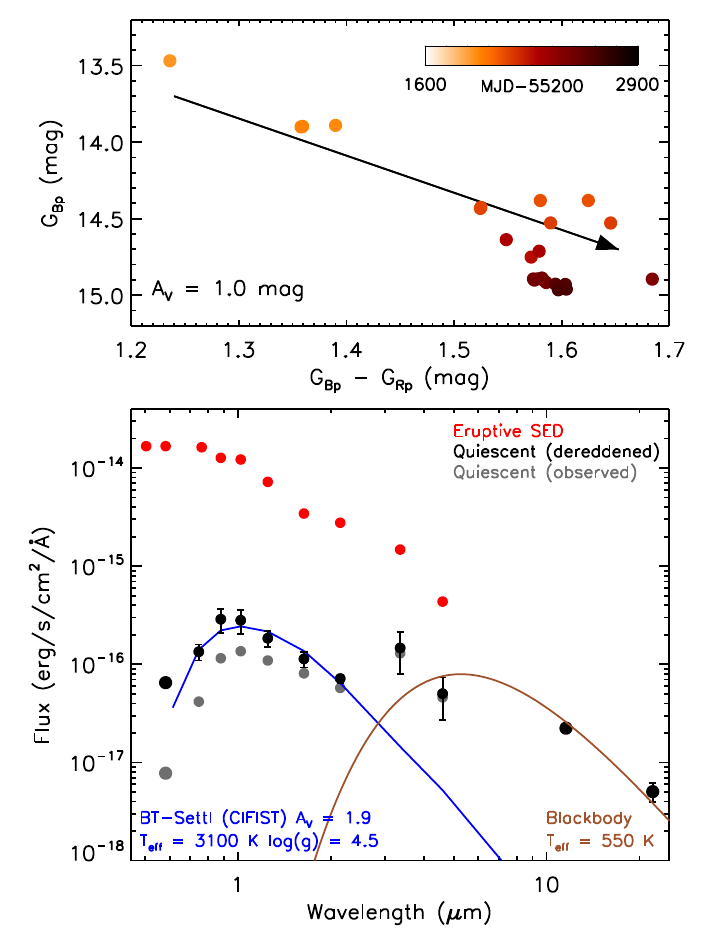}
    \caption{{\it Left}: Multi-wavelength light curves of L222\_78, colour coded by photometric bands. Error bars less than 0.1~mag are not shown. Open symbols are synthetic magnitudes computed from the SED best-fitting atmospheric model and spectra. Vertical dashed lines mark the observation dates of two spectral epochs. The ending point of the rising stage is shown by the dotted line (around MJD 57080/2015-02-27). {\it Upper Right}: Gaia colour-magnitude diagram with an extinction vector ($A_V = 1.0$ mag). The data points are colour-coded by the observation time. {\it Lower Right}: The pre-outbursting (grey), dereddened (black) and eruptive (red) SEDs. A BT-Settl model (blue) is fit to the pre-outbursting SED and a blackbody model (brown) is fit to the IR excess beyond 2 $\mu$m.} 
    \label{fig:lc}
\end{figure*}


Two spectra of L222\_78 were observed in 2021 and 2023. The XSHOOTER spectra \citep{Vernet2011} were obtained on the Very Large Telescope on April 16th 2021. We used optical and NIR arms of the XSHOOTER with slit widths of 0.9$"$ in the optical and 0.6$"$ in the NIR, with exposure times of 210s and 300s, respectively. The data was reduced using the pipeline package built on the {\sc reflex} platform \citep{Freudling2013}. We applied the {\sc molecfit} software to correct the telluric absorption and to generate wavelength solution of 2nd-order polynomial \citep{Smette2015, Kausch2015}. On May 9th 2023, we obtained NIR spectra from the Folded-port Infra-Red Echellette (FIRE) spectrograph on the Magellan Baade Telescope \citep{Simcoe2013}. The spectra are composed of four 90s exposures in the ABBA mode using the 0.6$"$ slit. An A0-type telluric standard is obtained afterwards. We applied the {\sc firehose v2.0} pipeline to reduce the data \citep{gagne2015}, using similar methods described in \citet{Guo2020}. 

\section{Results}
\label{sec: result}
\subsection{Photometric features}
\label{sec:photometry}

The optical to infrared lightcurves of L222\_78, from 2010 to 2023, are presented in Figure~\ref{fig:lc}. We divide the entire eruptive event into four stages based on the light curve morphology: the pre-eruptive pre-outbursting stage, the rising stage, the post-peak decaying stage and the re-brightening stage. The $K_s$ (VVV and SOFI) and {\it WISE} photometry are obtained throughout all stages, and the post-eruptive decaying stage is well-covered by {\it Gaia} light curves. Several sparse multi-colour photometry ($g$, $r$, $i$ and {\it VISTA} filters) are obtained at both pre-outbursting and post-eruptive stages, which are applied to measure the spectral energy distribution (SED) of the target. 


The rising stage of the outburst started in Oct 2013 and reached the peak brightness before Feb 2015, captured by the {\it WISE} and {\it Gaia} data. In the optical, the outburst has an exceptionally fast eruptive stage, as $\Delta G > 6.3$~mag within 170 days, resembling the behaviour of FU Ori. The first {\it Gaia} epoch was 338 days after the beginning of the near-infrared outburst. Nevertheless, a lower limit of the pre-outbursting $G$~magnitude ($G_{\rm low}$) is estimated by the photometric transformation equation as 20.17~mag\footnote{According to the {\it Gaia} EDR3 document, $G - g < -(g - i)+1.0$}, and a higher value is found from the synthetic photometry of the best fitting photospheric model (see \ref{sec:sed}). The duration of the eruptive stage across different bands will be further addressed by analytical functions in \S\ref{sec:rising}. 

After reaching the photometric maximum, the Gaia light curves of L222\_78 entered a decaying stage, whilst its infrared brightness remained relatively constant. We present the $G_{Bp}$ and $G_{Rp}$ colour-magnitude diagram in Figure~\ref{fig:lc}. The optical decay agrees with the extinction law assuming $R_V = 3.1$ and $A_V = 1.0$~mag \citep{WangS2019}. Based on the {\it Gaia} light curves, we generated a time-extinction correlation to reconstruct the peak brightness in VVV filters. The variation of extinction is not unusual on FUOrs, which has been referred to as the dust grain condensation when the wind collides with the envelope \citep[see][]{Siwak2023} or the dust lifted by the outflow during the ejection outburst. Such a dimming event is observed on V1057~Cyg, which faded 2 mag in 3 years after reaching the peak \citep{Herbig1977eruptive}.  The post-peak $A_V$ enhancement has also been confirmed in several FUOrs, including V960~Mon and Gaia21bty \citep{Hackstein2015, Siwak2023}. 

A low-amplitude colourless brightening trend is detected in the {\it ATLAS} light curves (see Figure \ref{fig:lc} and the online supplementary file), which is inconsistent with the extinction law mentioned above.  The same rising trend is detected in the mid-infrared, indicating a piling up of warm material in the circumstellar disc, which resembles the behaviour of PGIR 20dci \citep{Hillenbrand2021}. However, in the latter case, there was a secondary outburst a decade after the primary outburst, which has not happened yet on L222\_78. Nevertheless, we observe a change in the slope of the NIR continuum spectra, which can be interpreted as variable extinction (see \S\ref{sec:spec}). Therefore, we conclude that the recent rising trend on L222\_78 is likely attributed to a mixture of increasing mass accretion rate and clearing of line-of-sight extinction. The enhancement in the mid-infrared brightness also suggests the existence of warm dust in the circumstellar disc.  


\begin{figure}
\centering
\includegraphics[width=3.1in]{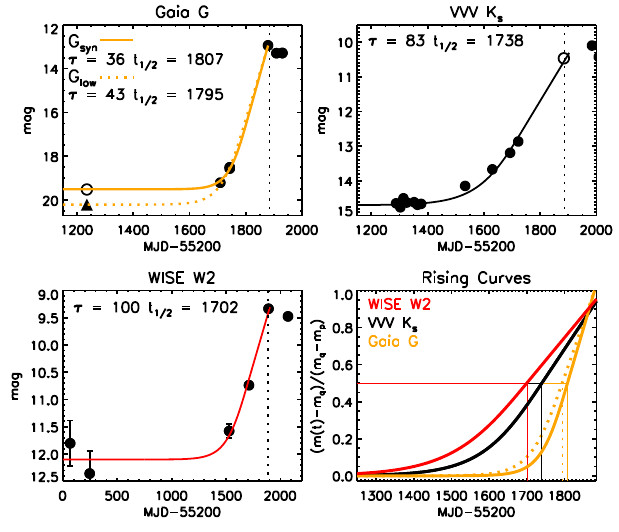}
    \caption{Rising light curves and analytical rising functions of L222\_78. Synthetic brightness (open circles) and lower limit (triangle) in $G$ are presented. The normalised curves are shown in the bottom right with $t_{1/2}$ marked out. The ending of the raising stage is marked by the dashed vertical lines.}
    \label{fig:rising_lc}
\end{figure}

\subsection{SED and bolometric luminosity}
\label{sec:sed}
We present the pre-outburst and outbursting SEDs of L222\_78 using photometric data obtained between 0.5 to 22 $\mu$m. The pre-outbursting SED comprises VVV photometry taken in 2010, the {\it VPHAS+} $i$-band in 2013 and {\it WISE} bands from our PSF photometry. We assume no significant variability ($\Delta m > 0.5$~mag) during the pre-outbursting stage. For the in-outburst (peak) brightness, we adopted {\it Gaia} and {\it NEOWISE} detection at/near the photometric maximum. The VVV~magnitudes were measured $\sim$$100$ days after the photometric maximum, which was affected by the enhancement of the extinction. Hence, the peak VVV magnitudes were estimated by correcting the extinction estimated from the {\it Gaia} light curves (see \S\ref{sec:photometry}). We then performed a least-$\chi^2$ fitting of the pre-outbursting SED to BT-Settl models \citep{Barber2006, Caffau2011, Allard2011} using the VOSA tool \citep[][]{Bayo2008} with a fixed Gaia distance (1.08 kpc) and $A_V$ ranging between 0 to 5 mag.

The pre-outbursting SED fitting result suggests that the progenitor has an effective temperature of 3100 K (1$\sigma$ confidence: 2769 - 3373 K) and a bolometric luminosity, $L_{\rm bol} = 0.16 \pm 0.02$~$L_\odot$. The stellar luminosity is a lower limit as there was barely any photospheric emission having been detected. The star could be heavily embedded and therefore be intrinsically bluer and more massive. The infrared excess can be fit by a single temperature blackbody (550$\pm$25 K), although the fitting around 3.6 $\mu$m is poor, suggesting a temperature gradient in the circumstellar disc. From the best fitting photosphere to the pre-outbursting SED, we compute the synthetic magnitude in $G$ and $r$ during the pre-outbursting stage, as $G_{\rm syn} = 19.5$~mag and $r_{\rm syn} = 21.4$~mag. Accurate measurements of the $L_{\rm bol}$ at the outbursting stage rely on the precise estimation of $A_V$. The post-outbursting spectra suggest very low extinction, with $A_V < 1$~mag (see \S\ref{sec:spec}). Therefore, the outbursting bolometric luminosity is between 9 to 16~$L_\odot$ when assuming $A_V$ ranging between 0 to 1~mag. Under the assumption that the accretion luminosity ($L_{\rm acc}$) is roughly equal to the bolometric luminosity during the eruptive stage, we estimated the peak mass accretion rate ($\dot{M}_{\rm acc}$) is 0.8 to 1.4 $\times$ 10$^{-5}$ $M_\odot$ yr$^{-1}$, by simply applying an approximated correlation, $\dot{M}_{\rm acc} = L_{\rm acc}R_*/(GM_*)$, \citep{Gullbring1998}. The peak absolute brightness of this source ($M_G \sim 3$ mag) indicates is considerably fainter than any observed stellar merger  \citep[$M_V < -3$ mag,][]{Karambelkar2023} but not inconsistent with a star-planet merger, such as the candidate event ZTF~SLRN-2020 \citep{De2023}. However, such events only last for a few hundred days. In contrast, L222\_78 has been in the outbursting phase for nearly a decade, suggesting a larger reservoir of the accreted material (i.e. the young accretion disc).


\begin{figure*}
\centering
\includegraphics[width=5.2in]{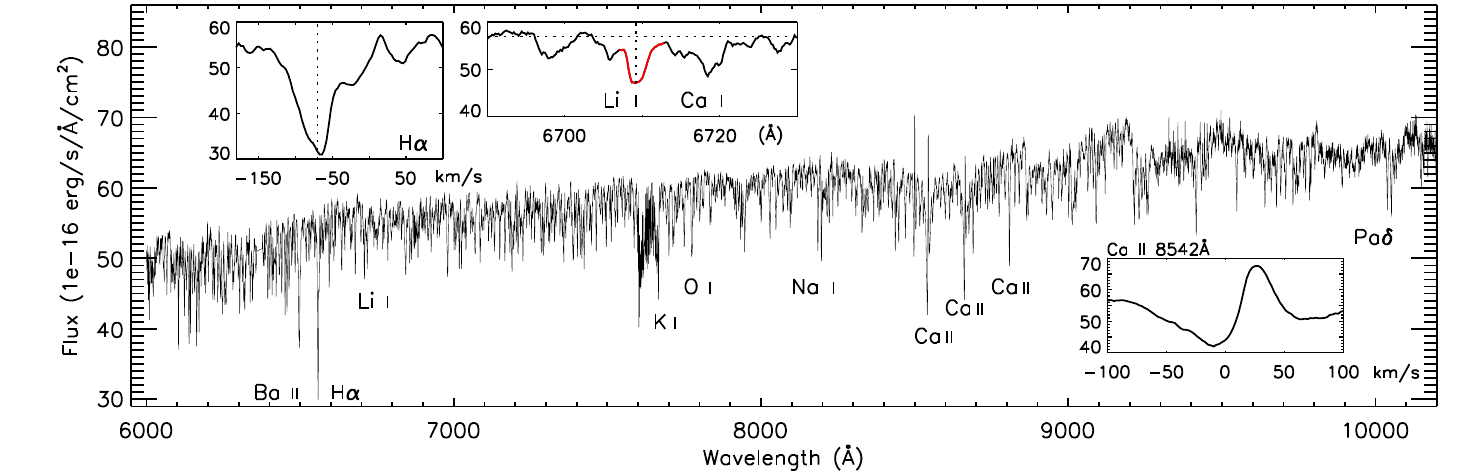}
\includegraphics[width=5.2in]{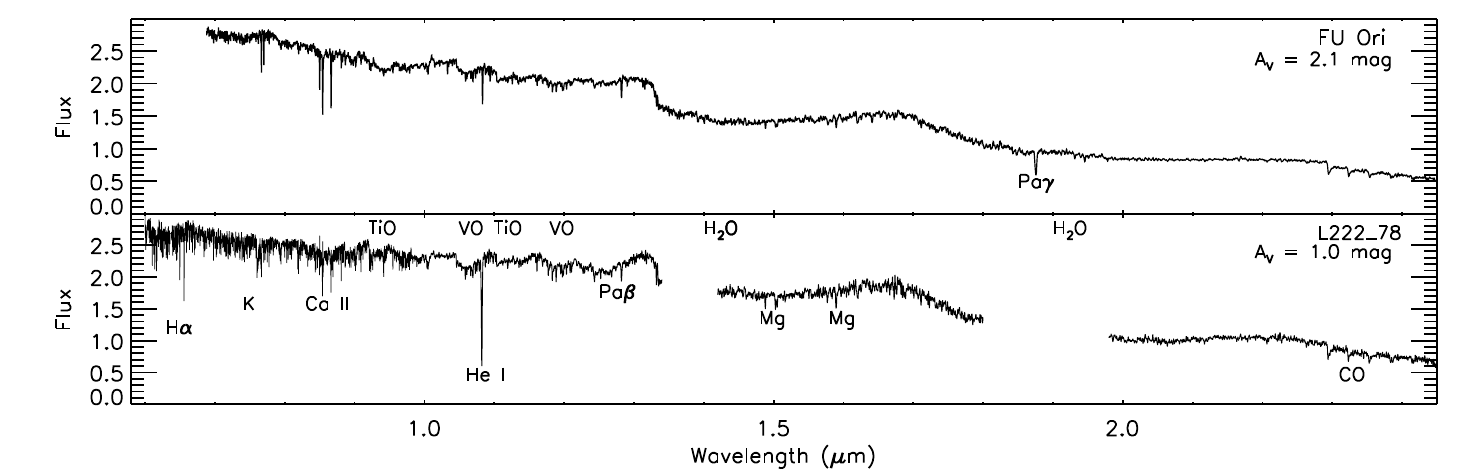}
    \caption{{\it Upper:} Optical spectrum of L222\_78, with line profiles of H$\alpha$, Li {\sc i} and Ca {\sc ii} absorption lines. {\it Lower}: Dereddened NIR spectra of FU~Ori \citep[IRTF;][]{Connelley2018} and L222\_78 (XSHOOTER). Spectral features are marked individually on the plot.}
    \label{fig:compare_spec}
\end{figure*}

\subsection{Rising Timescale}
\label{sec:rising}

The rising stage of L222\_78 was captured by a broad wavelength range, including an exceptional optical amplitude ($\Delta G > 6.3$~mag) which places it as one of the highest amplitude eruptive events on YSOs. We applied analytical functions to describe the exponential rising stage in $G$, $K_s$, and $W2$ using observed and synthetic~magnitudes. We adopted the two-step formalism originally designed by LSG23,
\begin{align}
t < t_{1/2}:\,\,\,\,\,\,\,\,\, &m(t) = m_q - \frac{m_q - m_p}{1+e^{-(t-t_{1/2})/{\tau}}} \\
t \geq t_{1/2}:\,\,\,\,\,\,\,\,\, &m(t) = m_q - (m_q - m_p)(0.5 + 0.5(t-t_{1/2})/{2\tau})
\end{align}
where $t_{1/2}$ is the time when the brightness is enhanced by half of the amplitude and $\tau$ is a timescale parameter. Additionally, $m_q$ and $m_p$ are the pre-outbursting and peak~magnitudes obtained from the time series which are treated as constants in our fitting. As designed, the photometric maximum ($m_p$) is reached at $t_{1/2} + 2\tau$, where the rising stage ends. This formalism reflects the classical rising morphology of most YSO outbursts in LSG23, composed of a slow-rising stage at the beginning that gradually accelerates before reaching a constant rate of brightening until reaching the photometric maximum.

Here, we used the photometric maximum captured by {\it Gaia} and {\it NEOWISE} near MJD~57083, and the synthetic peak~magnitude in $K_s$ (measured from the SED fitting above) as $m_p$. For $m_q$, we took the mean~magnitude in $K_s$ and $W2$ before MJD~56400, plus the $G_{\rm syn}$ and $G_{\rm lim}$ estimated in previous sections. We applied the $\chi^2$ minimization method on two free parameters ($t_{1/2}$ and $\tau$) with grid sizes of 1 day. The light curves, best-fitting analytical curves, and fitting parameters are presented in Figure~\ref{fig:rising_lc}. In most cases, the analytical functions fit the data well, except in $K_s$ where the brightness increased slightly faster than the model in the first 200 days.


To compare the rising stages across different wavelengths, we present the normalised rising curves on the bottom right panel of Figure~\ref{fig:rising_lc}. We found that the choice of pre-outbursting magnitude, either $G_{\rm syn}$ or $G_{\rm lim}$, does not significantly affect the fitting results in $G$. The infrared bands ($W2$ and $K_s$) started rising earlier and reached the $t_{1/2}$ point 60 to 100 days quicker than the $G$-band. The curve in $G$ is much steeper (smaller $\tau$) than the infrared, indicating a faster-rising nature in optical, with 170 days in the rising stage. However, this rising timescale is much longer than is typically the case in stellar mergers, though there are exceptions \citep[][and references therein]{Karambelkar2023}, and they often fade quickly afterwards \citep[][]{Tylenda2011}. The delay between the optical and infrared rising light curves agrees with the phenomenon predicted by the outside-in propagation of the FUOr outburst \citep{Cleaver2023}. However, the time delay observed on L222\_78 is five times shorter than the delays on Gaia17bpi and Gaia18dvy \citep[about 500 days][]{Hillenbrand2018, Szegedi-Elek2020}. Theoretically, the duration of the delay is consistent with the viscous timescale of the disc, proportional to the square root of the disc mass, which should be similar among low-mass YSOs. We will discuss the triggering mechanisms in \S\ref{sec: discussion}.




\subsection{Spectroscopic features}
\label{sec:spec}

 During a FUOr-type event, the viscously heated inner accretion disc becomes more luminous than the stellar photosphere, hence a cool but bright spectrum is observed in the infrared, and a hotter G/K-type spectrum in the optical. The optical spectrum of L222\_78, obtained after the outburst, is presented in the upper panel of Figure~\ref{fig:compare_spec}. It contains a forest of absorption features that are very similar to other FUOrs \citep[e.g.][]{Hillenbrand2023}, rather than other eruptive events, such as stellar mergers. The blueshifted $H\alpha$ absorption feature and the P Cygni profiles on Ca {\sc ii} indicate a wind launched during the outburst, a characteristic property of FUOrs \citep[see][]{Hillenbrand2019, Szabo2021}. We detected a broad Li $\lambda$6708\AA$\,$ absorption line (EW = $0.57 \pm 0.08$ \AA, integrated between 6707 to 6713\AA), as a common indicator of a young age \citep[e.g.][]{Bayo2011, Campbell-White2023}, and well matches the 1-2 Myr age estimation (Class II YSO). Additionally, the Na {\sc i} lines exhibit broad spectral profiles (FWHM = 86 km/s), which indicates a possible disc origination and is consistent with the Keplerian velocity at the 0.02 AU radius of a 0.2 $M_\odot$ star. Such a bright gas disc has been predicted by theoretical models \citep[][]{Liu2022} and has been observed on several young eruptive YSOs that were classified as bona fide FUOrs \citep[][]{Park2020, Szabo2022}. More detailed spectral features are presented in the online supplementary material.

 In the infrared, the unique spectral absorption features of FUOrs often resemble cool stars, such as the water vapour absorption arising from a low-gravity disc atmosphere, H {\sc i} lines, deep He {\sc i} absorption at 1.08 $\mu$m (typically associated with strong non-collimated winds), CO bandheads at 2.3 $\mu$m and many other molecular bands \citep[e.g. VO, TiO;][]{Connelley2018}. Notably, no jet/outflow feature (e.g. [Fe {\sc ii}], H$_2$) is detected. In the lower panel of Figure~\ref{fig:compare_spec}, we compared the dereddened XSHOOTER spectrum of L222\_78 with the spectrum of FU~Ori published in \citet[][]{Connelley2018}. Both spectra exhibit similar features, although He{\sc i} absorption is much deeper on L222\_78. Following the method applied in \citet[][]{Connelley2018}, we estimated the line-of-sight extinction of L222\_78 by aligning the $H$-bandpass continuum with FU~Ori, as $A_V = 1.0$~mag. By fitting the CO overtone models to the spectrum beyond 2.3 $\mu$m \citep[see details in][]{Contreras2017b}, we derived the radial velocity of the CO absorption feature as $-10 \pm 5$ km/s, with an effective temperature of $T_{\rm eff }= 3500 \pm 500$ K.
 

Spectroscopic variability is detected between the two epochs and interpreted as variable extinction, see a comparison between the two NIR spectra of L222\_78 in the online supplementary file. The difference in the spectral slope between the two epochs can be well-fitted by slightly modifying the line-of-sight extinction by $A_V=1.2$~mag. We measured $A_H = $ 0.16~mag between the two spectra, which is smaller than the contemporary photometric variability ($\Delta H$ = 0.24~mag).  The dereddened spectra of L222\_78 show great similarity to the dereddened IRTF spectra of FU Ori, assuming $A_V = 2.1$~mag \citep{Liu2022}. The low extinction estimated from our NIR spectra suggests that L222\_78 should have recovered from the post-peak extinction event. Compared with the estimated extinction in the pre-outburst SED ($A_V = 1.9$~mag), it is quite reasonable to assume that the outburst has cleared out the immediate line-of-sight extinction, also contributing to the high optical amplitudes.


\subsection{L222\_78 as a young stellar object}

We provide the following evidence to support a pre-MS classification of L222\_78. First, this target has been classified as a member of a young open cluster VdBH 221 according to the {\it Gaia} data with a membership probability of unity \citep[see the catalogue in][and Fig. C1 ]{Dias2021}.
We also find a few cold gas clumps located in the vicinity of this target, such as PGCC G354.05+02.95 \citep[10\arcmin away,][]{Planck2016}, although L222\_78 is not directly projected into this cold clump. Second, the pre-outburst SED models suggest $T_{\rm eff}\sim$3100 K and $L_{\rm bol}\sim0.16L_\odot$, placing this target in the zone of pre-MS stars on the H-R diagram, with an infrared excess. According to evolutionary models \citep{Baraffe2015}, this target is a $\sim$0.2 $M_\odot$ YSO with an age of 1-2 Myr and radius of 1.1 $R_\odot$. The 1-2 Myr of age is younger than the 10 Myr age of the VdBH 211 cluster \citep[][]{Dias2021}, although multi-generation star formation is common among young clusters. Additionally, the pre-outburst NIR colour indices ($J - H = 0.71$ and $H - K_s =$ 0.43) are consistent with the intrinsic colours of Class II YSOs \citep{Meyer1997}. The infrared spectral index $\alpha$ \citep[][measured from 2 to 24 $\mu$m]{Lada1987}, suggests L222\_78 is a Class II YSO ($\alpha=$-1.1). The optical spectrum of L222\_78 is very similar to previously detected FUOrs, such as the blue-shifted H$\alpha$ and P Cygni profiles on the Ca {\sc ii} absorption lines. Finally, we detected the Li absorption line, a well-established indicator of a young age (see \S\ref{sec:spec}).


\section{Discussion, Remarks and Summary}
\label{sec: discussion}

This study presents an ongoing {  eruptive} object located in the VdBH 221 open cluster, with a 6.3~mag outburst observed in the Gaia time series with an optical rising stage of 170 days. A short time delay ($\Delta t_{1/2} = 100$ days) is observed between the infrared and optical rising stages, suggesting an outside-in propagated instability that originated in the circumstellar disc at a radius less than 0.1 AU.  {  Some lower amplitude variations are detected after the photometric maximum, attributed to the gradual changing of line-of-sight extinction and overall mass accretion rate. The prevalence of absorption features in the optical to NIR spectra agrees with the FUOr-type outburst. We conclude that this source is a bona fide FUOr-type object based on the current evidence and the criteria proposed in \citet{Connelley2018}. Long-term monitoring is required to determine the full timescale of this event and to establish a definitive classification.}

In the scenario of magneto-rotational activation at the dead zone \citep[][]{Armitage2001, Elbakyan2021, NayakshinElbakyan2023}, the burst is thermally triggered when the mid-plane temperature exceeds the critical temperature of $\sim$800~K. The disc at this location becomes ionised, and the heatwave propagates inwards from beyond 0.2 AU. However, the predicted time delay is on the scale of $10^3$~days, much longer than the delay observed on L222\_78. Nevertheless, for a short time delay, the instability must originate closer to the star \citep[i.e. 0.07 AU;][]{Liu2022}, which is a small radius for MRI activation. As a reference, such events were predicted to happen in the magnetic field dead zone located around  1 AU for a solar mass star \citep[see][]{NayakshinElbakyan2023}.

Alternatively, an outside-in burst could be introduced by an embedded massive proto-planet, which introduces the thermal instability bursts by opening a gap and piling up material behind its orbit \citep[TIP+EE,][]{Lodato2004, Nayakshin2023, NayakshinElbakyan2023}. In this scenario, the outburst starts outside the planetary orbit ($<$ 0.1 AU) and propagates inwards, creating a months-long time delay between the optical and infrared bands, consistent with the delay observed on L222\_78. The predicted peak mass accretion rate is on the order of $10^{-5}$ $M_\odot$ yr$^{-1}$, similar to our measurements on L222\_78. During the outbursting stage, the temperature of the surrounding proto-planetary material will become higher than the planet's temperature. As a result, the protoplanet can go through the “extreme evaporation” (EE) phase, during which it is disrupted thermally \citep[][]{Nayakshin2023}. 

After the initial outburst, a rise in the visual extinction ($A_V \sim 1$~mag) has been detected by the Gaia time series. This change in extinction could be caused by temporary circumstellar dust structures that form subsequent to the outburst, lifted by the strong wind launched from the ejection burst, consequently affecting the optical depth at shorter wavelengths. This extinction event has not changed the overall eruptive phenomenon observed in the Gaia light curves ($\Delta G > 6.3$ mag). This is unlike the rapid fading V1647 Ori-type objects, whose 5 mag outburst has completely faded in 1000 days. According to the most recent NIR spectra, the extinction has been cleared out in the past years.

We discover a continuous brightening trend in recent years, from optical (1 mag) to mid-infrared (0.8 mag), {  which is a novel feature compared to the traditional FUOr-type. This brightening trend can be explained as an increased emission of the system,} likely attributed to the eruptive energy heating up the circumstellar disc. Theoretically, a secondary rise is predicted by both the aforementioned TIP+EE and MRI models, attributed to the feeding from the evaporated planetary atmosphere or the change in the mean molecular weight of the gas as Hydrogen is ionised \citep[see details in][]{NayakshinElbakyan2023}.

Using VOSA we fit a grid of BT-Settl models to the pre-outbursting SED. The synthetic magnitudes derived from the best-fit model are $G_{\rm syn} = 19.5$ and $r_{\rm syn} = 21.4$ mag. Compared with the peak brightness, we find $\Delta G = 6.6$~mag and $\Delta r = 8.6$~mag, ranking L222\_78 among the highest amplitude outbursts on YSOs. The clearing of extinction during the outburst could partially contribute to the optical amplitude. The SED fitting result shows that L222\_78 is a low-mass YSO (0.2~M$_\odot$) with {  low} extinction ($A_V = 1.9$~mag). The peak mass accretion rate is estimated as up to  1.4 $\times$ 10$^{-5}$ $M_\odot$ yr$^{-1}$, indicating $\sim40$ $M_\oplus$ have been accreted since the beginning of the outburst. Remarkably, only a few FUOrs were detected within a comparable mass range in the literature, such as V2775~Ori and V346~Nor \citep{Caratti2011, Kospal2021}. Both sources have relatively heavy and gravitational-unstable circumstellar discs ($M_d > 0.25 M_\star$), making them good candidates for GI-triggered outbursts. Future sub-mm observations are anticipated to measure the disc mass of L222\_78, to unveil the underlying triggering mechanism. Currently, with a low foreground extinction, L222\_78 (despite its distance) is an excellent target for high-resolution follow-up observation, to enable studies on the outbursting behaviours of FUOrs.

\section*{Data Availability}
The WISE data underlying this article are publicly available at the IRSA server \url{https://irsa.ipac.caltech.edu/Missions/wise.html}. The SOFI, VVV and VST data are publicly available at the ESO archive \url{http://archive.eso.org/cms.html}. The VIRAC2$\beta$ version of the VVV/VVVX light curves has not yet been publicly released but is available on request to the first author. The raw photometric data obtained by REM and SMART are available upon request. The Gaia light curves and photometric measurements are published on the ESA website (\url{https://www.cosmos.esa.int/gaia}). The ATLAS data is available via the webpage of the survey. Reduced spectra are provided at \url{http://star.herts.ac.uk/~pwl/Lucas/GuoZ/VVVspec/}. 

\section*{Acknowledgements}

We thank Prof M. Connelley for sharing the spectrum of FU~Orionis originally taken by IRTF. We appreciate the instructive comments and suggestions from the anonymous referee. ZG is supported by the ANID FONDECYT Postdoctoral program No. 3220029. ZG acknowledges support by ANID, -- Millennium Science Initiative Program -- NCN19\_171. ZG, PWL, and CJM acknowledge support by STFC Consolidated Grants ST/R00905/1, ST/M001008/1 and ST/J001333/1 and the STFC PATT-linked grant ST/L001403/1. This work has made use of the University of Hertfordshire's high-performance computing facility (\url{http://uhhpc.herts.ac.uk}).

J.A.-G., J.B. and R.K. thank the support from ANID's Millennium Science Initiative ICN12\_009, awarded to the Millennium Institute of Astrophysics (MAS). J.A.-G. acknowledges support also from Fondecyt Regular 1201490. ACG acknowledges support from INAF-GOG "NAOMY: NIR-dark Accretion Outbursts in Massive Young stellar objects" and PRIN 2022 20228JPA3A - PATH. V.E. acknowledges the support of the Ministry of Science and Higher Education of the Russian Federation (State assignment in the field of scientific activity 2023, GZ0110/23-10-IF).
 D.M. gratefully acknowledges support from the ANID BASAL projects ACE210002 and FB210003, from Fondecyt Project No. 1220724, and from CNPq Brasil Project 350104/2022-0

We gratefully acknowledge data from the ESO Public Survey program ID 179.B-2002 taken with the VISTA telescope, and products from the Cambridge Astronomical Survey Unit (CASU). This work contains data from ESO programs 105.20CJ and 109.233U, using SOFI on NTT and XSHOOTER on VLT. This research has made use of the NASA/IPAC Infrared Science Archive, which is funded by the National Aeronautics and Space Administration and operated by the California Institute of Technology. This work has made use of data from the European Space Agency (ESA) mission {\it Gaia} (\url{https://www.cosmos.esa.int/gaia}), processed by the {\it Gaia}
Data Processing and Analysis Consortium (DPAC,
\url{https://www.cosmos.esa.int/web/gaia/dpac/consortium}). Funding for the DPAC has been provided by national institutions, in particular, the institutions participating in the {\it Gaia} Multilateral Agreement. This paper includes data gathered with the 6.5 meter Magellan Telescopes located at Las Campanas Observatory, Chile. We acknowledge the support from the 1.0 m SMART telescope (operated by the SMARTS Consortium). This publication makes use of VOSA, developed under the Spanish Virtual Observatory project supported by the Spanish MICINN through grant AyA2008-02156.

\label{lastpage}

\bibliographystyle{mnras}
\bibliography{reference} 



\appendix
\section{Photometric measurements}
\begin{figure*}
\includegraphics[width=5in]{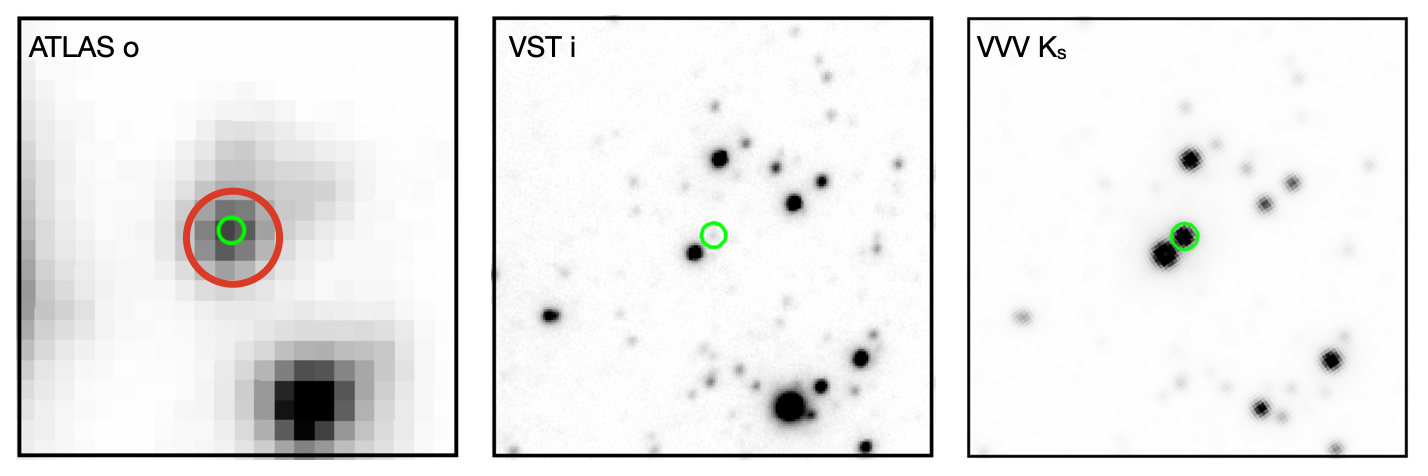}
    \caption{Photometric images of L222\_78 from  {\it ATLAS},  {\it VST} and the VVV survey. The size of the images is $40" \times 40"$. The location of L222\_78 is marked by the green circles with a radius of $1.2"$. The observation dates of the images are MJD 58007, MJD 56441 and MJD 57185, respectively. A larger aperture (5$"$, red) is placed on  {\it ATLAS} and  {\it REM} images to include both L222\_78 and its companion.}
    \label{fig:image}
\end{figure*}

\begin{figure*}
\includegraphics[height=2.in]{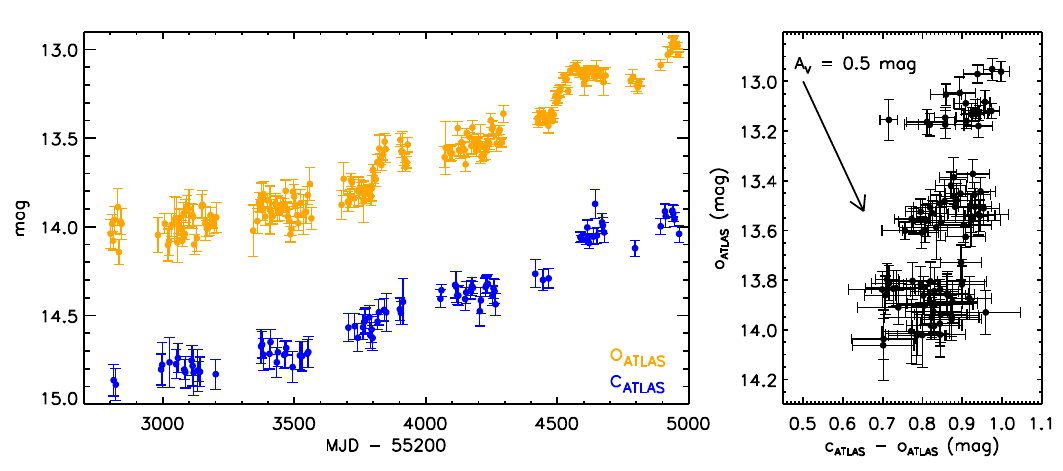}
\includegraphics[height=2.in]{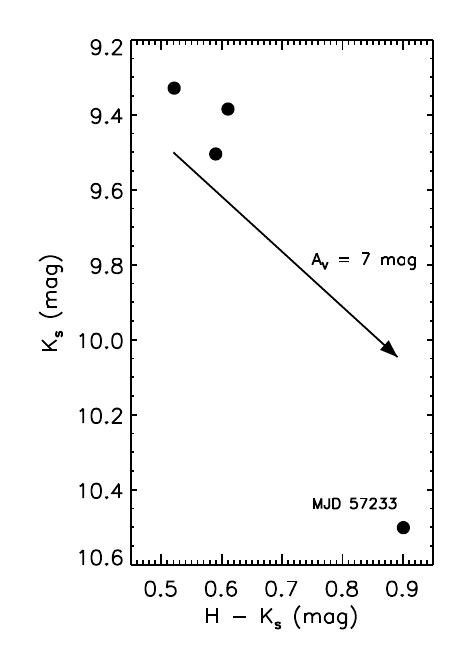}

    \caption{{\it Left}: {\it ATLAS} light curves of L222\_78 during the post-decaying rising stage ($o_{\rm ATLAS}$: yellow, $c_{\rm ATLAS}$: blue).   {\it Middle}: Colour magnitude diagram (CMD) of the {\it ATLAS} photometry. An extinction vector is presented with $A_V = 0.5$ mag. {\it Right}: Post-peak CMD of near-infrared bands. The data taken during the recent brightening stage is presented on the upper left corner.}
    \label{fig:ATLAS}
\end{figure*}

In this section, we provide more information on the aperture photometry methods that were applied to extract the brightness of L222\_78 from the raw images. Specifically, L222\_78 was observed by {\it VST} but does not appear in the {\it VPHAS+} DR2 catalogue, hence we obtained the reduced images via the ESO archive. As mentioned in the main text, a companion is located $2.2"$ away from L222\_78, which was brighter than our target during the pre-outbursting stage. In high spatial resolution images, e.g. from {\it VST} and {\it VISTA}, these two sources are well-distinguished. However, custom-written programs are necessary for low-resolution images where the two targets are blended. For the same reason, we downloaded more than 1700 reduced images from the {\it ATLAS} forced-photometry server (pixel size $1.86"$), to obtain more reliable measurements than the pipeline-produced PSF photometry. Example images observed by {\it VISTA}, {\it VST} and {\it ATLAS} are presented in Figure \ref{fig:image}. 

We first extract the brightness of L222\_78 from the images with high spatial resolutions ({\it VST} and {\it SMARTS}), using aperture photometry with an aperture size of $1.5"$. Three bright and non-saturated nearby sources are selected as reference stars, with their brightness obtained from the {\it ATLAS} all-sky stellar reference catalogue \citep{Tonry2018a}. We used the standard deviation of the background and the statistical noise to calculate the photometric errors. We also measured the brightness of the companion source, from both {\it VST} and {\it SMARTS} images, as $g = 22.03 \pm 0.60$ mag, $r = 18.40 \pm 0.03$ mag, $i = 15.60 \pm 0.02$ mag and $z = 13.53 \pm 0.02$ mag. In addition, we used the functions from \citet{Tonry2018} to convert $g$, $r$, $i$-band magnitudes to {\it ATLAS} bands.

Before reducing the data from {\it REM} and {\it ATLAS}, we co-added individual images into 1-day bins, to increase the signal-to-noise ratio. On average, there are four to five images per day per bandpass. We then extracted the combined brightness of L222\_78 and the companion using a larger aperture (see Figure \ref{fig:image}). 
Finally, to obtain the brightness of L222\_78, we subtracted the expected brightness of the companion (measured above) from the combined brightness. The light curves and colour-magnitude diagrams of the {\it ATLAS} data are presented in Figure \ref{fig:ATLAS}. An overall colour-less rising morphology is seen between both bands, plus some short-timescale variations are detected in the $o_{\rm ATLAS}$-band. The colour variation disagrees with the extinction vector from \citet{WangS2019}.

\section{Spectroscopic features}

\begin{figure*}
\includegraphics[height=2.3in]{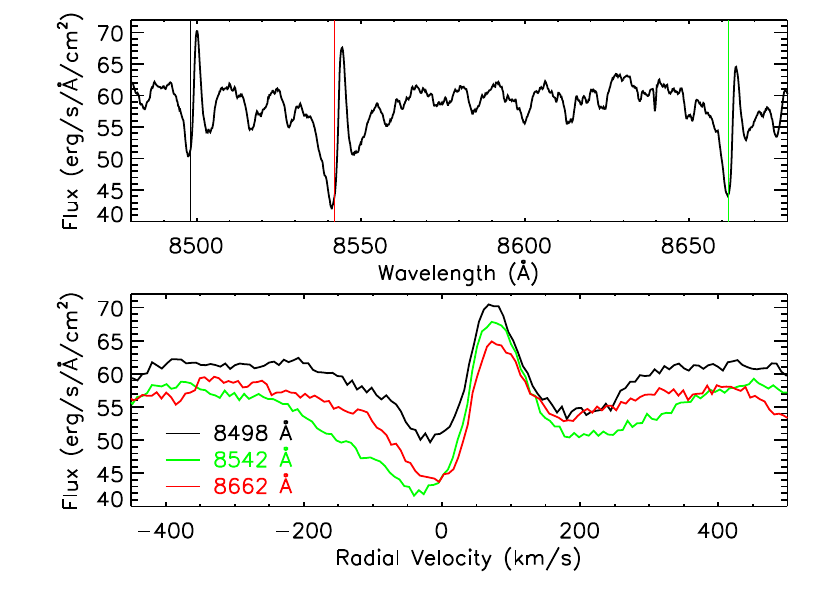}
\includegraphics[height=2.3in]{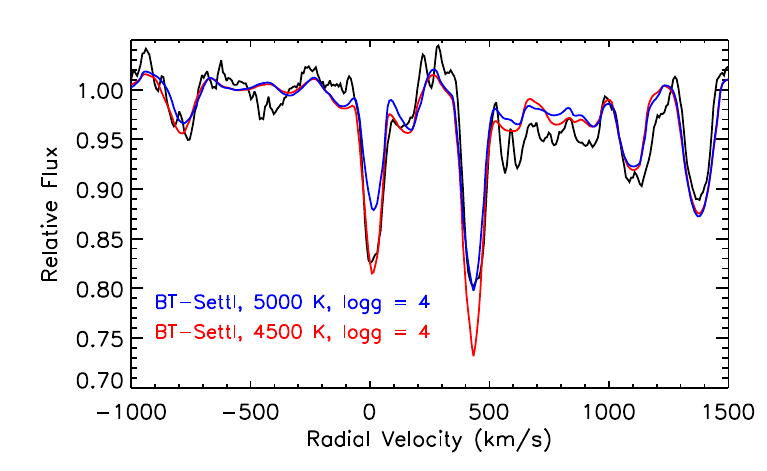}
    \caption{{\it Left}: Line profiles of three Ca {\sc ii} lines of L222\_78. {\it Right}: The Na {\sc i} doublet around 8200 \AA$\,$ (black), with theoretical spectra from BT-Settl models.}
    \label{fig:spec_features}
\end{figure*}

\begin{figure}
\includegraphics[width=3in]{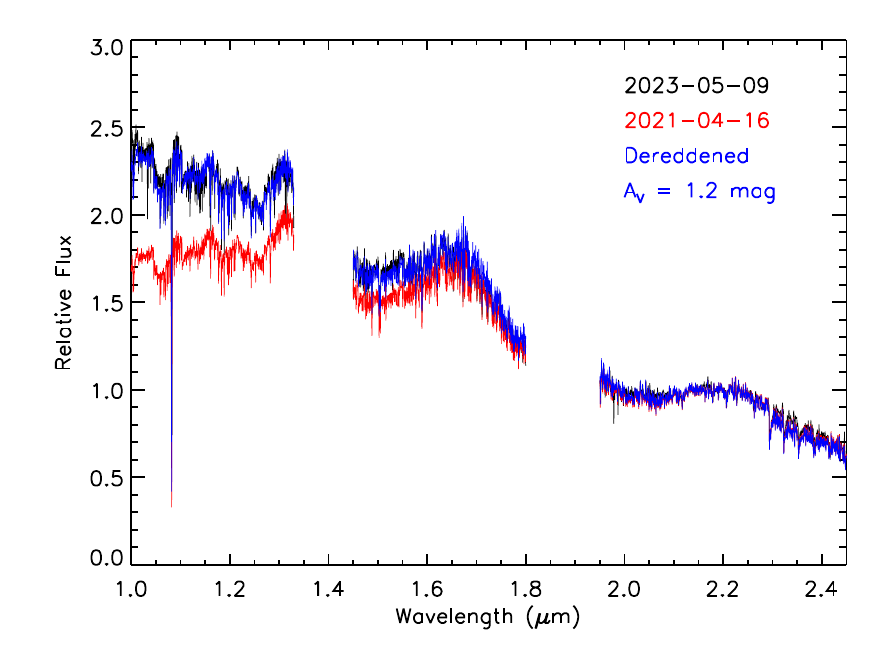}
    \caption{spectra of L222\_78 taken by FIRE (2023-05-09) and XSHOOTER (2021-04-16) spectrographs. The XSHOOTER spectrum is dereddened (blue) by $A_V = 1.2$~mag to match the continuum in the 2023 epoch.}
    \label{fig:NIR_continnum}
\end{figure}

Here we present more details on the spectral feature of L222\_78, including the P Cygni profiles on the Ca {\sc ii} triplet lines, indicating a strong wind launched by the outburst. We highlighted the Na {\sc i} absorption lines around 8190 \AA$\,$ in Figure \ref{fig:spec_features}, with synthetic spectra from the BT-Settl model \citep{Allard2011}. The model spectra are resampled using Gaussian profiles to match the resolution of the observed spectrum. The model spectrum is then broadened by a rotational velocity, assuming an edge-on viewing angle and using the convolution kernel calculated from the {\sc lsf\_rotate} program in {\sc idl}. The temperature (4500 - 5000 K), surface gravity (log g = 4.0), and rotational broadening (90 km/s) are selected by fitting our observed spectrum to synthetic models with the minimum $\chi$-square method. The rotational velocity fits a Kelperian disc located at 0.02 AU around a 0.2 $M_\odot$ star. The rotational broadening velocity also matches the FWHM of the line profile (86 km/s). According to theoretical models \citep{Liu2022}, such viscously heated gas discs can reach the order of 4000 K during a FUOr-type outburst. However, we note here that we can not completely rule out the possibility that the Na {\sc i} absorptions are raised from a fast spin stellar surface, although it is somewhat an extreme case. The two near-infrared spectra, from FIRE and XSHOOTER, are presented in Figure \ref{fig:NIR_continnum}. The difference between the two continuum spectra can be corrected by de-reddening the XSHOOTER spectrum with $A_V = 1.2$ mag.

\section{the young cluster VdBH 221}
\begin{figure*}
\includegraphics[width=2.8in]{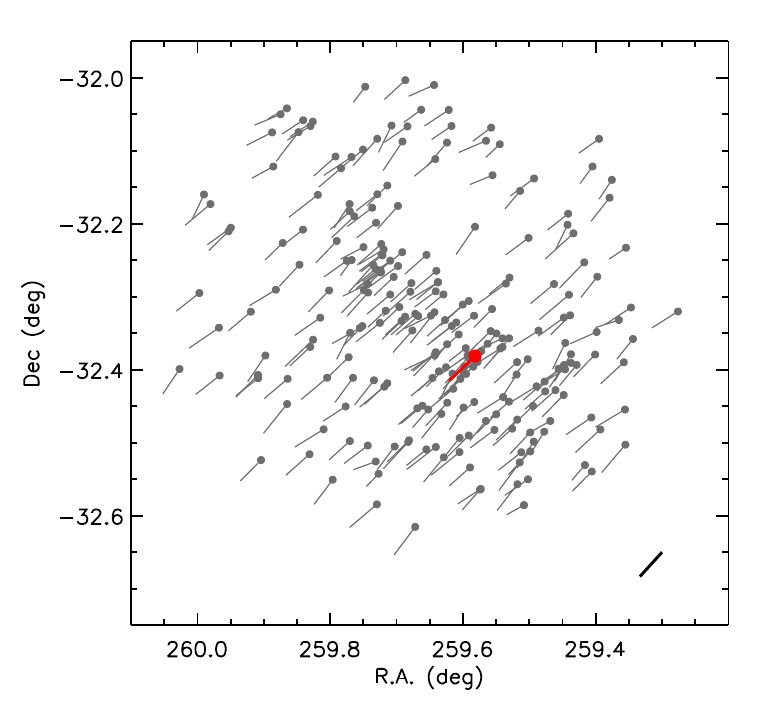}
\includegraphics[width=2.8in]{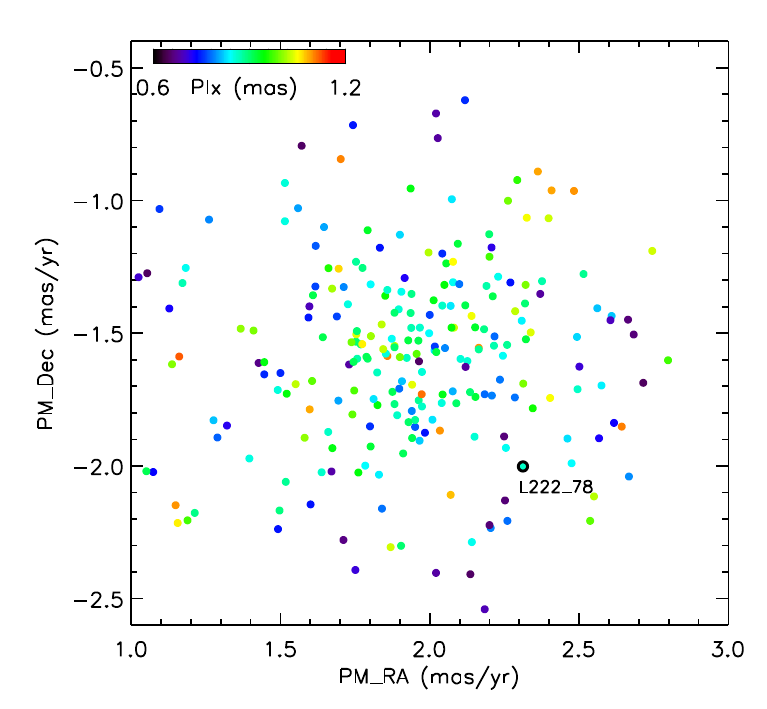}
    \caption{{\it Left}: The coordinates (in deg) of cluster members of VdBH 221, adopted from \citep{Dias2021}. L222\_78 is marked with a red colour. The {\it Gaia} proper motions are presented by the solid lines. The bottom right corner shows a scale of proper motions (pmRA = 2 mas/yr and pmDec = -2 mas/yr). {\it Right}: The proper motions of the cluster members of VdBH 221, including L222\_78, colour-coded by parallax in mas.}
\label{fig:gaia}
\end{figure*}

In this section, we present the evidence to support that L222\_78 is a member of the young open cluster VdBH 221. The {\it Gaia} coordinates and proper motions \citep[adapted from][]{Dias2021} are shown in Figure \ref{fig:gaia}. L222\_78 is located at 4.2 arcmin to the projected centre of the cluster, within the radius of 50\% of the cluster members ($\sim$ 8.6 arcmin). The VdBH 221 cluster has a median distance of 1045 pc (parallax = 0.86 mas, 1$\sigma$ = 0.12 mas), which is in line with the {\it Gaia} distance of L222\_78, as 1080 pc (parallax = 0.92$\pm$0.02 mas). The proper motions of L222\_78 (pmRA = 2.312, pmDec = -2.002) are also consistent with the median proper motions of the cluster.

\section{Photometric data}
\begin{table}
\caption{Photometric data on L222\_78}
\renewcommand\arraystretch{1.4}
\begin{tabular}{c c c c c c}
\hline
\hline
Date (MJD) & Band & Telescope & mag & err (mag) & method \\
\hline
56909 & $G$ & Gaia & 19.20 & 0.03 & $c$ \\
57078 & $G$ & Gaia & 12.94 & 0.01 & $c$ \\
57078 & $G_{\rm BP}$ & Gaia & 13.47 & 0.01 & $c$ \\
57078 & $G_{\rm RP}$ & Gaia & 12.23 & 0.01 & $c$ \\
56774 & $g$ & VST & 22.00 & 0.50 & $p$ \\
56441 & $r$ & VST & 20.83 & 0.60 & $p$ \\
56441 & $i$ & VST & 19.17 & 0.20 & $p$ \\
58001 & $o_{\rm ATLAS}$ & ATLAS &  14.04 & 0.08 & $p$ \\
58013 & $c_{\rm ATLAS}$ & ATLAS &  14.86 & 0.09 & $p$ \\
55408 & $Y$ & VISTA & 16.64 & 0.28 & $p$ \\
55408 & $Z$ & VISTA & 17.63 & 0.41 & $p$ \\
... & ... & ... & ... & ... & ... \\

\hline
\hline
\end{tabular}
\flushleft{The full version of this table can be found in the online supplementary file.\\
Method: $c$: data obtained directly from the catalogue. $p$: data reduced in this paper.}
\label{tab:data}
\end{table} 

\end{document}